\documentclass[twocolumn,superscriptaddress,showpacs,floatfix,amsmath,amssymb]{revtex4-1}

\usepackage{graphicx}
\usepackage[dvipdfmx,colorlinks=true,linkcolor=blue,citecolor=blue,urlcolor=blue]{hyperref}
\usepackage[utf8]{inputenc}
\usepackage{float}

\newcommand{\singlet}{{}^1\mathrm{S}_0}
\newcommand{\triplet}{{}^3\mathrm{P}_2}
\newcommand{\bLi}{${}^7$Li}
\newcommand{\fLi}{${}^6$Li}
\newcommand{\bYb}{${}^{174}$Yb}
\newcommand{\fYb}{${}^{173}$Yb}
\newcommand{\Er}{E_R^{\mathrm{Yb}}}

\newcommand{\nm}{\mathrm{nm}}
\newcommand{\msec}{\mathrm{ms}}
\newcommand{\s}{\mathrm{s}}

\newcommand{\mG}{\mathrm{mG}}
\newcommand{\Hz}{\mathrm{Hz}}
\newcommand{\kHz}{\mathrm{kHz}}
\newcommand{\MHz}{\mathrm{MHz}}

\begin{document}

\title{Experimental realization of ultracold Yb-\bLi\ mixtures in mixed dimensions}

\author{F.~Sch\"{a}fer}
\email{schaefer@scphys.kyoto-u.ac.jp}
\affiliation{Department of Physics, Graduate School of Science, Kyoto University, Kyoto 606-8502, Japan}

\author{N.~Mizukami}
\affiliation{Department of Physics, Faculty of Science, Kyoto University, Kyoto 606-8502, Japan}

\author{P.~Yu}
\affiliation{Department of Physics, Graduate School of Science, Kyoto University, Kyoto 606-8502, Japan}
\affiliation{Department of Physics, Harvard University, Cambridge, MA 02138, USA}

\author{S.~Koibuchi}
\affiliation{Department of Physics, Graduate School of Science, Kyoto University, Kyoto 606-8502, Japan}

\author{A.~Bouscal}
\affiliation{D\'{e}partement de Physique, \'{E}cole Normale Sup\'{e}rieure, PSL Research University, 24 rue Lhomond, 75005 Paris, France}

\author{Y.~Takahashi}
\affiliation{Department of Physics, Graduate School of Science, Kyoto University, Kyoto 606-8502, Japan}

\date{\today}

\begin{abstract}
	We report on the experimental realization of ultracold \bYb-\bLi\
	(Boson-Boson) and \fYb-\bLi\ (Fermion-Boson) mixtures. They are loaded into
	three dimensional (3D) or one dimensional (1D) optical lattices that are
	species-selectively deep for the heavy Ytterbium (Yb) and shallow for the
	light bosonic Lithium (Li) component, realizing novel mixed dimensional
	systems. In the 1D optical lattice the band structure of \fYb\ is
	reconstructed in the presence of \bLi. Spectroscopic measurements of the
	\bYb-\bLi\ mixture in the 3D lattice give access to the \bYb\ Mott-insulator
	structure. Ground state inter-species scattering lengths are determined to
	be $|a_{\rm bg}({}^{174}{\rm Yb}$-${}^{7}{\rm Li})|=(1.11 \pm 0.17)~\nm$ and
	$|a_{\rm bg}({}^{173}{\rm Yb}$-${}^{7}{\rm Li})|=(1.16 \pm 0.18)~\nm$. The
	formation and characterization of an ultracold \fYb-\bLi\ mixture is a first
	step towards a possible realization of a topological $p_x + i\,p_y$
	superfluid in this system.
\end{abstract}

\maketitle

\section{Introduction}
\label{sec:intro}

In recent years, the strong and interdisciplinary effort towards the
realization of topological phases of matter has evolved, particularly bringing
topological insulators~\cite{hasan_colloquium:_2010} and
superconductors~\cite{qi_topological_2011} into focus. Here, chiral $p_x +
i\,p_y$ superconductors in two dimensions (2D) are of particular interest, as
the appearance of robust Majorana modes might play an important role in
realizing fault-tolerant quantum computation~\cite{nayak_non-abelian_2008}.
Currently, Majorana modes are discussed in 1D nanowires (see
e.g.~\cite{mourik_signatures_2012, albrecht_exponential_2016}) and other
solid-state systems; for example, ${\rm Sr}_2{\rm RuO}_4$ is shown to form a
$p$-wave superconductor~\cite{kallin_chiral_2012}. Fine-tuned control of the
possible Majorana modes in these systems seems difficult, and it is highly
desirable to introduce additional platforms to realize topological superfluids
with good control of their properties. Ultracold atom systems could be
promising candidates for this application~\cite{goldman_topological_2016}.
Additionally, using multi-species systems is a familiar pathway for broadening
the range of accessible physical questions and, in particular, realizing
mixed-dimensional systems~\cite{nishida_universal_2008}.

Following these ideas, it was shown as early as ten years ago that a
two-species Fermi gas with one species confined in a 2D plane and immersed
into a 3D Fermi sea of the other species could lead to $p_x + i\,p_y$
superfluidity facilitated by inter-species $s$-wave
interaction~\cite{nishida_induced_2009}. Later, using
Berezinskii-Kosterlitz-Thouless theory, it was found that Fermi-Bose mixtures
in mixed dimensions also support this topological
superfluid~\cite{wu_topological_2016}. More detailed calculations revealed
that in considering higher-order contributions, the $p$-wave gap can be even
larger than previously expected~\cite{caracanhas_fermibose_2017}. The authors
of~\cite{caracanhas_fermibose_2017} also provide detailed calculations of the
transition behavior of the two-species fermionic Ytterbium (\fYb) and bosonic
Lithium (\bLi) system. Assuming suitable inter-species scattering lengths,
critical temperatures on the order of $0.07\, T_{\rm F}$, with $T_{\rm F}$ the
Fermi temperature, are predicted. These limits are certainly challenging.
However, they are not too low for state-of-the-art
experiments~\cite{navon_equation_2010}. Indeed, the present work reports on
the realization and characterization of a quantum degenerate, mixed
dimensional \fYb-\bLi\ system. 

Much effort has been spent in understanding the interactions in the sister
system of bosonic \bYb\ and fermionic \fLi~\cite{dowd_magnetic_2015,
schafer_spectroscopic_2017}, where even the realization of a double superfluid
was reported~\cite{roy_two-element_2017}. In this context, our work strives to
expand on these efforts by reversing the roles of the particles, switching out
bosonic for fermionic Yb and fermionic for bosonic Li. To complete the
picture, we also report on a quantum degenerate \bYb-\bLi\ mixture. 

The present paper is organized as follows. In Sec.~\ref{sec:expt}, we
introduce the experimental setup and details for reaching double quantum
degeneracy. Section~\ref{sec:results} summarizes our first characterization of
the mixtures obtained, that is, the measurement of the inter-species
scattering lengths (Sec.~\ref{sec:thermalization}), the creation of a mixed
dimensional system (Sec.~\ref{sec:modulationspectroscopy}) and the
high-resolution spectroscopy of the mixture in a 3D optical lattice
(Sec.~\ref{sec:3P0spectroscopy}). A final discussion of the results in
Sec.~\ref{sec:discussion} concludes our work.

\section{Experiment}
\label{sec:expt}

The experiment proceeds along similar lines as our earlier reported
works~\cite{konishi_collisional_2016}. In brief, a hot atomic beam is formed
starting from a dual species oven heated to about $350~^\circ {\rm C}$, which
contains Yb and Li in natural abundances as well as enriched \fLi.  The
isotope shifts of the optical transition frequencies are, for Yb, in the few
GHz range and, for Li, in the 10 GHz range. Thus, isotope-selective slowing
and cooling of both species is possible with only slight adjustments of the
cooling laser frequencies. We first slow down Yb in a Zeeman slower operating
on the strong $\singlet\,$-$^1{\rm P}_1$ transition followed by a
magneto-optical trap (MOT) on the narrow $\singlet\,$-$^3{\rm P}_1$
intercombination line. In a second step, Li is slowed and trapped relying on
the D2 transition. As a comment, we note that the hyperfine splitting for the
\bLi($^2{\rm S}_{1/2}$) ground state is, at $803.5~\MHz$, much larger than
that for \fLi\ at $228.2~\MHz$. Therefore, the inclusion of a $^2{\rm
S}_{1/2}(F = 1)$-$^2{\rm P}_{3/2}$ repumper laser to the standard $^2{\rm
S}_{1/2}(F = 2)$-$^2{\rm P}_{3/2}$ Zeeman slowing light is crucially important
for slowing sufficient numbers of \bLi\ atoms for loading into the MOT.
Compression of the MOT followed by reduction of MOT beam detunings and
intensities further cools the atomic clouds. This improves phase matching for
loading into our crossed far-off-resonance trap (FORT), where forced
evaporative (Yb) and sympathetic (Li) cooling are performed.

\begin{figure}[tb]
	\centering
	\includegraphics[width=7.5cm]{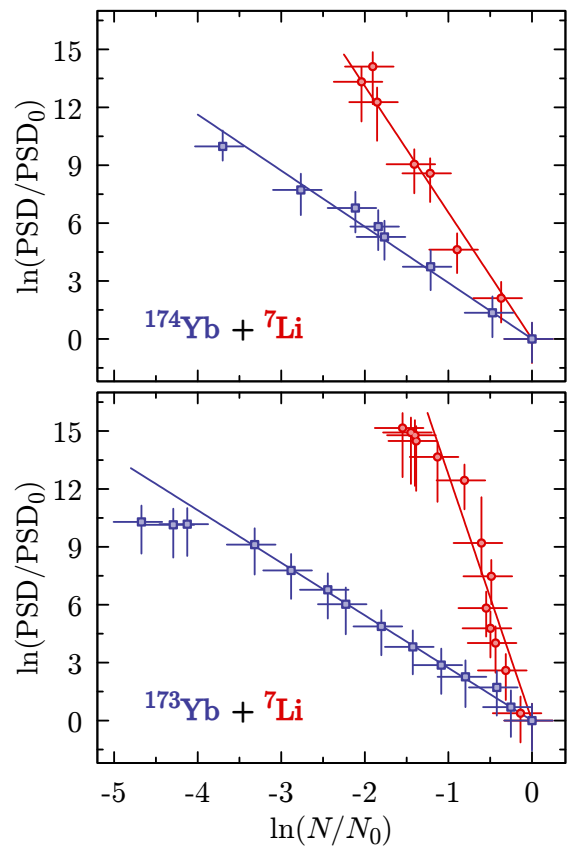}
	\caption{The path to double quantum degeneracy in mixtures of \bYb-\bLi\ 
		(upper panel) and \fYb-\bLi\ (lower panel). In each case, the development
		of the normalized phase-space density (PSD$/$PSD$_0$) and the normalized
		atom numbers ($N/N_0$) during forced evaporation and sympathetic cooling
		are shown as points. PSD$_0$ and $N_0$ are the respective values at the
		beginning of the evaporation ramp. Error bars account for uncertainties in
		the cloud and trap parameters. The trajectories for Yb (blue) and Li (red)
		can be roughly described and fitted by power laws (straight lines). See
	the main text for details on their interpretation.} 
	\label{fig:evap}
\end{figure}

For the case of a \bYb-\bLi\ mixture, we load the Yb and Li MOT for $14~\s$
and $0.5~\s$, respectively, which results in a typical value of $60\times10^5$
Yb atoms at $95~\mu{\rm K}$ and $1.5\times10^5$ Li atoms at $210~\mu{\rm K}$
in the crossed FORT at the beginning of the evaporation ramp. For \fYb-\bLi,
atom numbers after $20~\s$ and $0.3~\s$ of loading are $46\times10^5$ for Yb
at $80~\mu{\rm K}$ and $0.7\times10^5$ for Li at $115~\mu{\rm K}$. During
forced evaporation and sympathetic cooling, the crossed FORT lasers are
reduced from their initial powers to the final values in individually
optimized ramps within $7.65~\s$ ($9.8~\s$) for \bYb-\bLi\ (\fYb-\bLi). All
experiments reported in the present work were performed at low magnetic bias
fields, typically a few $100\,\mG$. Development of phase-space density (PSD)
and atom number ($N$) during evaporation normalized by their respective
initial values, PSD$_0$ and $N_0$, for both mixtures are shown in
Fig.~\ref{fig:evap}. From the double logarithmic representation, we see that
the initial stages of the evaporation are well described by a power law and we
extract $\gamma = -{\rm d\,ln} ({\rm PSD}/{\rm PSD}_0)/{\rm d\,ln} (N/N_0)$, a
measure for the evaporation efficiency~\cite{ketterle_evaporative_1996}, by
linear fit to the data. For the \bYb-\bLi\ mixture $\gamma_{\rm Yb} = 2.9(1)$
and $\gamma_{\rm Li} = 6.5(2)$ are obtained. Similarly, in the \fYb-\bLi\
evaporation sequence $\gamma_{\rm Yb} = 2.7(1)$ and $\gamma_{\rm Li} =
12.8(6)$ are observed. The cooling efficiencies for both Yb isotopes are
similar while sympathetic cooling of \bLi\ seems to proceed more effectively
in combination with \fYb. In the latter case, saturation of the \fYb\ PSD, in
particular, is also visible. The generally large $\gamma_{\rm Li}$ are a
result of the sympathetic cooling where the Li PSD is increased not at the
expense of the number of Li atoms, but at the cost of the coolant species, Yb.

\begin{figure}[tb]
	\centering
	\includegraphics[width=7.5cm]{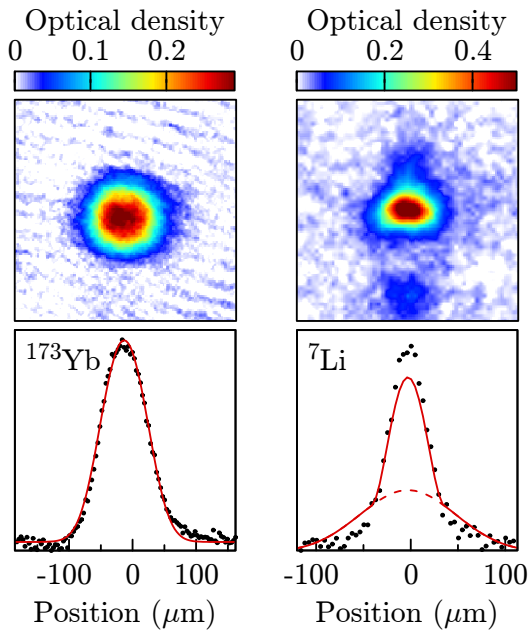}
	\caption{A quantum degenerate mixture of \fYb\ (left panels) and \bLi\ (right
		panels). The clouds have been imaged $15~\msec$ ($5~\msec$) after release
		from the trap for Yb (Li). The top panels show false-color representations
		of the obtained absorption images. The lower panels show projections of
		the data (points) and fit results (lines) on the horizontal axis.
		Fermionic \fYb\ was fitted by a Fermi gas distribution and bosonic \bLi\
		by a bimodal distribution (dashed line gives thermal component). The
		apparent discrepancy between fit and data seen in the Li projection is due
		to some Li atoms being ejected from the main cloud and giving rise to an
		additional contribution to the projection, see main text for further
		discussion.} 
	\label{fig:173Yb7Li}
\end{figure}

At the end of the evaporation ramp, the trap frequencies for both mixtures are
$(\omega_x,\omega_y,\omega_z) = 2\pi \times (38, 58, 221)~\Hz$ for Yb and
correspondingly $2\pi \times (250, 395, 1795)~\Hz$ for Li, where $z$ is in the
vertical direction. (The uncertainties of those values are on the order of
10\%.) The obtained quantum degenerate \fYb-\bLi\ mixture is shown in
Fig.~\ref{fig:173Yb7Li}. We have $N \approx 62\,000$ \fYb\ atoms at $T =
87~{\rm nK}$ and from the fugacity of the Fermi gas distribution, we find by
fit $T/T_{\rm F} \approx 0.4$, where $T_{\rm F}$ is the Fermi temperature. No
optical pumping techniques have been performed on \fYb\ during evaporation,
and we expect about equal populations of the six spin ground states. A precise
determination of the condensate fraction using an unrestricted bimodal fit for
the \bLi\ cloud is difficult. We therefore opted to fix the temperature to the
value obtained from the fit to the \fYb\ cloud and obtain $N_{\rm BEC} \approx
4\,200$ atoms in the Bose-Einstein condensate (BEC) and $N_{\rm th} \approx
7\,400$ atoms in the thermal component for \bLi. The time-of-flight absorption
image of \bLi, see left panel of Fig.~\ref{fig:173Yb7Li}, shows a slight
fragmentation of the atomic cloud, the origin of which is not yet understood.
By application of a magnetic field gradient after release of the atoms from
the trap, we determined that all Li atoms are actually spin polarized in the
$m_F = 0$ state and remaining field gradients cannot cause this splitting. We
therefore surmise that the probable cause is some unresolved dynamics occurring
as the crossed FORT is being turned off. Turning our attention to the
Boson-Boson \bYb-\bLi\ mixture (not shown), we report $N_{\rm BEC} \approx
76\,000$, $N_{\rm th} \approx 47\,000$, $T = 110~{\rm nK}$ for \bYb\ and
$N_{\rm BEC} \approx 12\,000$, $N_{\rm th} \approx 14\,000$ for \bLi\ again
assuming equal temperatures for both species.

\section{Analysis and results}
\label{sec:results}

Towards the long-term goal of forming a topological superfluid, three main
ingredients are of particular importance~\cite{caracanhas_fermibose_2017}: (i)
the creation of a Fermi-Boson quantum degenerate mixture at only few nK, (ii)
the formation of a mixed-dimensional system and (iii) information on the
inter-species $s$-wave scattering length. In the present initial survey, the
experimental setup was not designed to reach the required temperatures to
address the first point. Instead, we concentrate on the remaining points by
both determining the inter-species background scattering length,
Sec.~\ref{sec:thermalization}, and realizing and probing two different
mixed-dimensional systems, Secs.~\ref{sec:modulationspectroscopy}
and~\ref{sec:3P0spectroscopy}. 

\subsection{Inter-species scattering length}
\label{sec:thermalization}

\begin{figure}[tb]
	\centering
	\includegraphics[width=7.5cm]{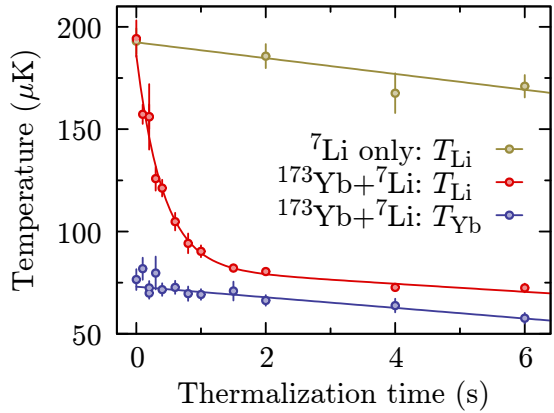}
	\caption{Overview of experimental data to determine the \fYb-\bLi\
		inter-species scattering length. After preparation of a
		temperature-imbalanced thermal mixture, the temperatures of Yb (blue
		points) and Li (red points) have been measured for thermalization times of
		up to $6~\s$. For reference, the temperature evolution of a Li-only sample
		has also been taken (yellow points). While the temperature of Li in the
		mixture has been fitted by an exponential decay (red line), the other data
		have been approximated by straight line fits (blue, yellow lines). For
		each thermalization time, the individual temperatures have been estimated
		from a series of measurements at different expansion times and the error
		bars correspond to the uncertainties in the temperature estimation from
		these data.}
	\label{fig:thermalization}
\end{figure}

Similar to earlier determinations of the \bYb-\fLi\ inter-species scattering
length~\cite{ivanov_sympathetic_2011, hara_quantum_2011}, we perform
cross-thermalization measurements in cold, but thermal mixture samples of
\bYb-\bLi\ and \fYb-\bLi\ mixtures. The experiment starts by loading either
mixture into the crossed FORT and confining the sample for $3~\s$ to reach a
steady state of $60$--$80~\mu{\rm K}$. Then, by slight power modulation of the
horizontal FORT laser beam at $9.5~\kHz$, the Li sample is selectively heated
to $150$--$200~\mu{\rm K}$. Through keeping the temperature-imbalanced mixture
in the trap for a variable time before releasing it, we gain access to the
temperatures by taking a series of absorption images at different expansion
times and the result for the \fYb-\bLi\ case is shown in
Fig.~\ref{fig:thermalization}. During the measured thermalization times, the
Yb temperature $T_{\rm Yb}$ is found to vary little, while the Li temperature
$T_{\rm Li}$ quickly approaches $T_{\rm Yb}$. This shows that Yb can be
treated as a sufficiently large heat bath at constant temperature. We
attribute the gap in final temperatures to a residual small miscalibration of
the two independent imaging systems. A control measurement in which Yb is
blasted by resonant $\singlet\,$-$^1{\rm P}_1$ light at zero holding time
confirms that all cooling of the Li is due to Yb.

Standard cross-thermalization analysis~\cite{ivanov_sympathetic_2011} then
gives access to the modulus of the inter-species scattering length. In the
error analysis uncertainties of the atom numbers by $20\%$, the temperatures
by $10\%$ and the densities by $30\%$ are considered. For the Fermi-Bose
mixture, $|a_{\rm bg}({}^{173}{\rm Yb}$-${}^{7}{\rm Li})|=(1.16 \pm 0.18)~\nm$
is obtained. In the corresponding experiment for the Bose-Bose mixture, we
find $|a_{\rm bg}({}^{174}{\rm Yb}$-${}^{7}{\rm Li})|=(1.11 \pm 0.17)~\nm$.
These values should be compared to previous calculations, albeit done for
different Li hyperfine states, where scattering lengths of $+1.80~\nm$ and
$+1.74~\nm$ have been reported~\cite{brue_magnetically_2012}. This shows that
while good order-of-magnitude agreement is achievable, the details of the
inter-species interaction potentials are quite challenging to model correctly.

\subsection{Band structure of \fYb\ in a 1D optical lattice}
\label{sec:modulationspectroscopy}

The aspired realization of a $p_x + i p_y$ superfluid heavily depends on the
formation of a novel mixed-dimensional system. Here, by means of a strong 1D
optical lattice, we realize an array of 2D \fYb\ fermionic systems in a 3D
\bLi\ bosonic bath. The 1D optical lattice is formed by two horizontally
counter-propagating laser beams~\cite{konishi_collisional_2016} with
wavelength $\lambda_{\rm L} = 532~\nm$. To reduce the impact of the
differential gravitational sag between Yb and Li in this and the following
experiments, a compensating beam at the same wavelength, focused slightly above
the atoms, is utilized~\cite{konishi_collisional_2016}. The lattice depth for
Yb is set to $15~\Er$, with $\Er = \hbar^2 k_{532}/(2m_{\rm Yb})$ being the Yb
recoil energy, where $m_{\rm Yb}$ is the Yb atomic mass and $k_{532} =
2\pi/\lambda_{\rm L}$. In this situation the Li lattice depth is only
$0.7~E_R^{\mathrm{Li}}$ which is too shallow to support a bound state in the
optical lattice, permitting the Li atoms to still freely move in 3D space. To
confirm formation of the mixed-dimensional system, we then perform lattice
modulation spectroscopy~\cite{heinze_multiband_2011} to reveal the Yb band
structure. After adiabatically ramping up the optical lattice, its depth is
modulated by about $5~\Er$ for $0.3~\msec$ and then ramped down to zero in
$0.2~\msec$ converting higher band excitations via band mapping to real
momenta that can be observed after time-of-flight absorption imaging. 

\begin{figure}[tb]
	\centering
	\includegraphics[width=7.5cm]{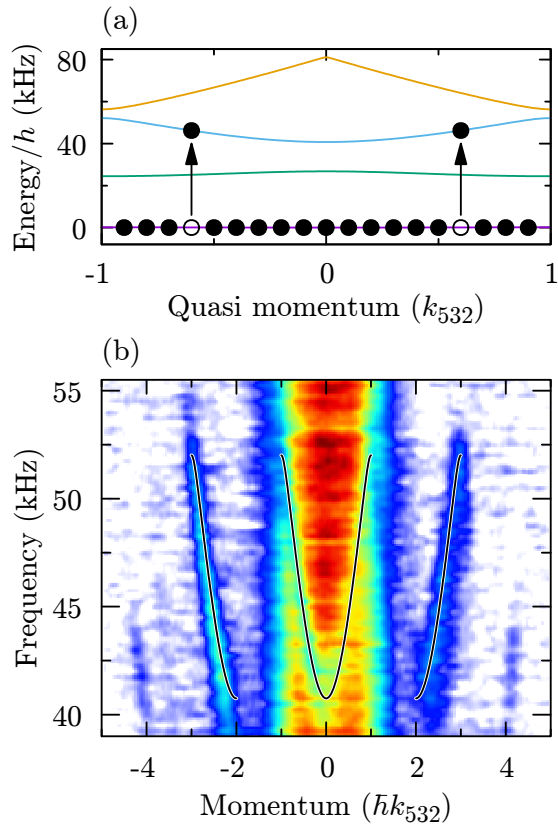}
	\caption{Lattice modulation spectroscopy of a double quantum degenerate
		\fYb-\bLi\ mixture in a deep 1D optical lattice. (a) Calculated first four
		bands of the lattice at $14~\Er$ lattice depth (purple to orange lines).
		The population of the first band by Yb Fermions (filled circles) and the
		excitation of a particle to the third band by lattice modulation (arrows),
		creating a hole in the first band population (open circle) are
		schematically indicated. (b) Experimental modulation spectroscopy data.
		The momentum distribution recorded by a band mapping procedure for
		different modulation frequencies reveals the creation of particle-hole
		pairs in the first and third lattice bands. The expected
		momentum-frequency dependence for this process is indicated (black lines).
		In the false-color representation blue (red) corresponds to a lower
		(higher) momentum population probability.}
	\label{fig:modulation}
\end{figure}

The process is schematically depicted in Fig.~\ref{fig:modulation}(a), where
the first four bands of the 1D lattice for Yb are shown. The Fermions
initially occupy the complete first band (lower dots) and modulation of the
lattice at the proper frequency leads to a particle-hole pair (filled and
empty circles) between predominantly the first and third band. Experimental
data is shown in Fig.~\ref{fig:modulation}(b), where the real momentum
distribution after time-of-flight is given versus the modulation frequency.
The formation of particle-hole pairs is visible as areas of
increased-decreased density. The black lines give the expected structure for a
lattice at $14~\Er$ which might indicate that the previously done calibration
of the lattice depth slightly overestimated the actual potential depth. Also
the weak signals at $\pm 4\, \hbar k_{\rm 532}$ could indicate imperfections
in the band mapping procedure. A repetition of the experiment with \bLi\
blasted away before ramping up the optical lattice (not shown) yields
identical results.

\subsection{Spectroscopy in a 3D optical lattice}
\label{sec:3P0spectroscopy}

\begin{figure}[tb]
	\centering
	\includegraphics[width=7.5cm]{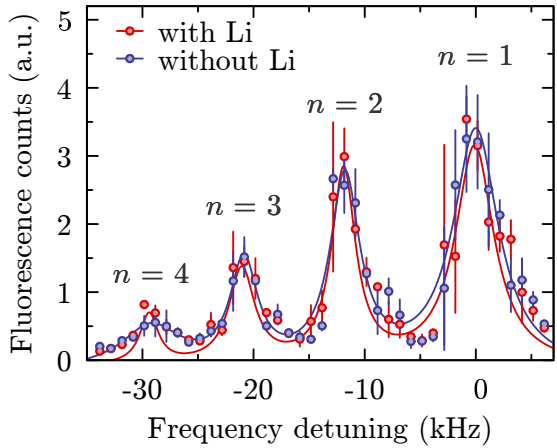}
	\caption{Measured $\singlet \rightarrow \triplet(m_J = 0)$ excitation
		spectrum (red points) of \bYb\ with \bLi\ in a 3D optical lattice at
		$15~\Er$. At this lattice depth, Li is not localized, while Yb forms a
		Mott-insulator state and atoms in lattice sites with different occupation
		numbers ($n = 1,2,3,4$) are separated due to interatomic interaction. For
		comparison, the experiment was repeated with the Li atoms removed before
		excitation (blue points). No significantly different excitation behavior
		is found. The lines are Lorentzian fits to the resonances.}
	\label{fig:spectroscopy}
\end{figure}

We now proceed to further reduce the dimensionality of Yb in the mixed
dimensional Yb-\bLi\ systems. This can be done in almost the same manner for
the \fYb-\bLi\ Fermi-Bose and the \bYb-\bLi\ Bose-Bose mixtures. For
experimental ease, we choose here the \bYb-\bLi\ system and load in into a 3D
cubic optical lattice at $15\,\Er$, where Yb forms a Mott-insulating state
while Li remains non-localized. In the case of \bYb, by exploiting the narrow
\bYb($\singlet \rightarrow \triplet$) transition, we are able to energetically
distinguish lattice sites with different Yb occupation numbers as previously
demonstrated for both a pure \bYb\ system~\cite{kato_laser_2016} and the
\bYb-\fLi\ mixture~\cite{konishi_collisional_2016,
schafer_spectroscopic_2017}. The results are summarized in
Fig.~\ref{fig:spectroscopy}, where a measurement in the presence of \bLi\
(red) is compared to a situation where the Li atoms were removed from the trap
just before loading the \bYb\ atoms into the lattice (blue). The well-resolved
resonances corresponding to occupation numbers up to $n = 4$ demonstrate the
successful formation of a Mott-insulating state in presence of a \bLi\ bosonic
background gas. We also note that possible shifts of the resonance frequencies
due to different interaction strengths of the Yb $\triplet$ and $\singlet$
states with Li have not been observed.

\section{Discussion and Conclusion}
\label{sec:discussion}

With the present set of experimental results, we demonstrate the realization
of doubly quantum degenerate mixtures of either \bYb\ or \fYb\ and \bLi. In
cross-thermalization measurements the background elastic scattering lengths
have been determined. They roughly agree with earlier theoretical
considerations and are generally, similar to the same mixtures involving \fLi,
quite small. By then loading the Fermi-Bose mixture into a 1D optical lattice,
a mixed dimensional regime has been achieved. As confirmation serve
measurements of the \fYb\ band structure and of the \bYb\ Mott-insulator
state. Thus, an important step towards topological $p_x + i\,p_y$ superfluids
has been taken. As expected, inter-species interaction effects are found to be
small and enhancement mechanisms such as suitable Feshbach resonances are
advantageous for reaching the required Fermi-Bose interactions. The current
setup strictly relies on sympathetic cooling to reach the Li quantum
degenerate regime. It is desirable to improve this situation by e.\,g.,
implementation of Li gray molasses cooling
techniques~\cite{burchianti_efficient_2014} and possibly the use of \bLi\
Feshbach resonances to reduce initial temperature and to enhance evaporation
efficiency. Finally, suitable lattice geometries that support the necessary
bosonic excitations, while remaining experimentally feasible to realize, need
to be explored.

The experiments detailed in the present work serve to establish additional
quantum degenerate mixtures in the toolbox of ultracold atomic physics. It is
the first large mass-imbalance Bose-Fermi system for the realization of mixed
dimensional geometries with a 3D bosonic background and a fermionic component
of reduced dimensionality.

\section*{Acknowledgments}
This work was supported by the Grant-in-Aid for Scientific Research of JSPS
Grants No.\ JP25220711, No.\ JP17H06138, No.\ 18H05405, and No.\ 18H05228, JST
CREST Grant No.\ JPMJCR1673 and the Impulsing Paradigm Change through
Disruptive Technologies (ImPACT) program by the Cabinet Office, Government of
Japan.

%

\end{document}